\documentclass[twocolumn,showpac,aps,prb,superscriptaddress]{revtex4-2}
\UseRawInputEncoding 
\usepackage{dcolumn}

\usepackage{amsmath}
\usepackage{amssymb}
\usepackage{bm}
\usepackage{txfonts}
\usepackage{mathdots}
\usepackage{graphicx}
\usepackage[T1]{fontenc}
\usepackage{xcolor}
\usepackage{float}
\usepackage{soul}

\begin{document}


 
\title{Correlation-driven origin of shallow electron pocket in Co$_{1/3}$TaS$_2$ revealed by ARPES and cluster perturbation theory}


\author{Wojciech Sas}
\affiliation{Institute of Physics, 10000 Zagreb, Croatia}
\affiliation{Institute of Nuclear Physics Polish Academy of Sciences, 31-342 Krak\'{o}w, Poland} 

\author{Yuki Utsumi Boucher}
\affiliation{Institute of Physics, 10000 Zagreb, Croatia}

\author{Seyed Ashkan Moghadam Ziabari}
\affiliation{Institute of Physics, 10000 Zagreb, Croatia}

\author{Gaurav Pransu}
\affiliation{Institute of Physics, 10000 Zagreb, Croatia}

\author{Trpimir Iv\v{s}i\'{c}}
\affiliation{Ru\dj{}er Bo\v{s}kovi\'{c} Institute, 10000 Zagreb, Croatia}

\author{Ivana Vobornik}
\affiliation{Istituto Officina dei Materiali (IOM)-CNR Area Science Park - Basovizza, I-34149 Trieste, Italy}

\author{Jun Fujii}
\affiliation{Istituto Officina dei Materiali (IOM)-CNR Area Science Park - Basovizza, I-34149 Trieste, Italy}

\author{Naveen Singh Dhami}
\altaffiliation[Current affiliation:]{Universit\'{e} Paris-Saclay, CNRS, Laboratoire de Physique des Solides, 91405 Orsay, France}
\affiliation{Institute of Physics, 10000 Zagreb, Croatia}

\author{Bruno Gudac}
\affiliation{Department of Physics, Faculty of Science, University of Zagreb, 10000 Zagreb, Croatia}

\author{Mario Novak}
\affiliation{Department of Physics, Faculty of Science, University of Zagreb, 10000 Zagreb, Croatia}

\author{L{\'a}szl{\'o} Forr{\'o}}
\affiliation{Department of Physics and Astronomy, University of Notre Dame, Notre Dame, IN 46556, USA}

\author{Neven Bari\v{s}i\'{c}}
\affiliation{Department of Physics, Faculty of Science, University of Zagreb, 10000 Zagreb, Croatia}
\affiliation{Institute of Solid State Physics, TU Wien, 1040 Vienna, Austria}

\author{Ivo Batisti\'{c}}
\email{ivo@phy.hr}
\affiliation{Department of Physics, Faculty of Science, University of Zagreb, 10000 Zagreb, Croatia}

\author{Petar Pop\v{c}evi\'{c}}
\email{ppopcevic@ifs.hr}
\affiliation{Institute of Physics, 10000 Zagreb, Croatia}

\begin{abstract}

We investigate the electronic structure and Fermi surface of Co$_{1/3}$TaS$_2$ using angle-resolved photoemission spectroscopy (ARPES) combined with theoretical modeling beyond standard density functional theory (DFT+U). A shallow electron pocket, the so-called $\beta$ feature, is observed at the Fermi level near the corner of the superlattice Brillouin zone, representing the first experimental observation of this feature in an intercalated TaS$_2$ compound. Similar pockets have been reported in $X_{1/3}$NbS$_2$ ($X$ = Co, Cr, Ni), where their surface versus bulk origin remains actively debated. Because conventional DFT+U does not capture this feature, we employ cluster perturbation theory (CPT) to incorporate an explicit treatment of strong electron correlations ($U$) on the Co sites. CPT successfully reproduces the $\beta$ feature, demonstrating its origin from correlation-driven bulk states rather than surface effects. To further substantiate this conclusion, we studied a reduced Co-content sample, Co$_{0.22}$TaS$_2$, where the reduced charge transfer modifies the Co-derived states near the Fermi level. Its electronic structure remains largely similar to that of pristine 2H-TaS$_2$, showing only a minor overall energy shift and lacking the $\beta$ feature, consistent with disrupted long-range Co ordering and modified orbital character near the Fermi level. We demonstrate that the $\beta$ feature arises from strong local correlations on the Co sites and requires long-range crystallographic order among intercalated Co atoms to maintain coherence. These results highlight the importance of strong electronic correlations in magnetically intercalated transition-metal dichalcogenides and provide a microscopic understanding of features not captured by conventional DFT+U.

\end{abstract}

\keywords{ARPES; Co intercalation; band structure; transition-metal dichalcogenides; resonance}

\maketitle


\section{Introduction}

Two-dimensional magnetic materials have attracted considerable attention for spintronic applications and the design of magnetic heterostructures \cite{Zhang2024}. Among them, transition metal dichalcogenides (TMDs) intercalated with magnetic atoms provide a versatile platform, where symmetry, magnetic interactions, and electron correlations intertwine and can be tuned by carrier concentration or external fields \cite{Wang2020, Witte2024, Hatanaka2023, Gao2024}.

Cobalt-intercalated TMDs have drawn particular interest owing to their rich topological magnetic phase space and the resulting anomalous transport phenomena. Large anomalous Hall effects (AHE) have been observed in both Co$_{1/3}$NbS$_2$ and Co$_{1/3}$TaS$_2$ \cite{Ghimire2018, tenasini2020, Mangelsen2021, Popcevic2022, Tanaka2022, Yang2022, Liu2022, Park2022, Takagi2023}, and attributed to Berry curvature contributions arising from the non-collinear magnetic order of these compounds. This topological character of the electronic structure manifests in a broad range of related effects: a strain-tunable anomalous Hall effect and a large spontaneous Nernst effect in Co$_{1/3}$NbS$_2$ \cite{Chen2025, Gu2025, Khanh2025}, electrical control of the topological magnetic state via ionic gating \cite{Kim2025}, and current switching of topological spin chirality together with a topological magneto-optical Kerr effect in Co$_x$TaS$_2$ \cite{Farhang2026}. Collectively, these findings establish Co-intercalated TMDs as promising candidates for functional spintronic and magnetoelectric device applications, and have stimulated intense interest in understanding the underlying electronic structure of Co$_{1/3}$TaS$_2$ \cite{Noh2026, Kar2026, Luo2026}.

In Co$_{1/3}$TaS$_2$, Co atoms occupy octahedral sites coordinated by S atoms and form a $\sqrt{3}\times\sqrt{3}$ superlattice rotated by 30$^\circ$ relative to the 1$\times$1 unit cell of 2H-TaS$_2$ \cite{vandenBerg1968, Parkin1980a, Parkin1980b, Parkin1983}. 
The system was long believed to order antiferromagnetically below $T_{\mathrm{N}} \approx 35$ K \cite{Parkin1980a, Parkin1980b}. However, recent studies have revealed a more complex magnetic phase diagram near $x \approx 1/3$, arising from strong magnetic frustration and a pronounced sensitivity to the Co concentration \cite{Park2024}.
For overdoped samples ($x \gtrsim 0.33$), the system exhibits coplanar helical antiferromagnetic order below $T_{\mathrm{N}} \approx 35$ K, whereas in underdoped samples ($x \lesssim 0.325$) two distinct antiferromagnetic transitions occur at $T_{\mathrm{N1}} \approx 38$ K and $T_{\mathrm{N2}} \approx 26.5$ K, including a noncoplanar triple-$\mathbf{q}$ configuration associated with a large AHE and finite scalar spin chirality \cite{Takagi2023, Park2023}. Slight deviations from the 1/3 concentration lift the frustration and favor one of the competing interactions, leading to qualitatively different magnetic and consequently transport properties, as well as other characteristics that are crucial for the functionality and application-relevant behavior of these materials.

Such pronounced changes in physical properties cannot be reconciled within a simple rigid-band picture, suggesting an active role of intercalant-derived electronic states in shaping the low-energy electronic structure. Indeed, angle-resolved photoemission spectroscopy (ARPES) studies of related intercalated compounds, Co$_{1/3}$NbS$_2$, Cr$_{1/3}$NbS$_2$, and Ni$_{1/3}$NbS$_2$, have revealed a shallow electron band near the Fermi level, which we refer to here as the "$\beta$ band" \cite{Sirica2016, Popcevic2022, Tanaka2022, Yang2022, Utsumi2024}. Its surface versus bulk origin remains debated, and its microscopic origin is unresolved. 
Conventional density functional theory (DFT) and its extension DFT+U reproduce the main bands of these intercalated TMDs, but fail to capture the $\beta$ band, indicating that correlation effects beyond the static mean-field description of DFT+U on the intercalant sites play a crucial role.

These observations raise a central question: what is the origin of the shallow $\beta$ band in intercalated TMDs, and how do electronic correlations and intercalant-derived states govern its presence and coherence?

In this work, we combine ARPES measurements on Co$_{1/3}$TaS$_2$ and underdoped Co$_{0.22}$TaS$_2$ with theoretical modeling beyond conventional DFT. Using cluster perturbation theory (CPT), we explicitly incorporate local electron correlations on the Co sites and analyze the shallow $\beta$ band in Co$_{1/3}$TaS$_2$. The underdoped sample allows us to examine how reduced cobalt concentration and the associated disorder in the intercalant lattice affect the coherence of Co-derived bands and modify the low-energy electronic states. These measurements, combined with CPT modeling, provide a framework for understanding how intercalant charge, orbital character, and correlations determine the low-energy electronic structure in magnetically intercalated TMDs.

\section{Experiment and Methods}

Single crystals of Co$_{1/3}$TaS$_2$ were grown by chemical vapor transport using iodine as the transport agent \cite{Friend1977}. The grown crystals were characterized by electrical resistivity, X-ray diffraction, energy dispersive X-ray spectroscopy (EDX), and magnetic susceptibility. EDX analysis indicated 32\% intercalation. To achieve this level, it was important to use $\sim$15\% more Co than the stoichiometric composition would suggest. In the following, this sample is referred to as Co$_{1/3}$TaS$_2$. Magnetic susceptibility revealed an antiferromagnetic transition at $T_\mathrm{N} = 37$~K. $T_\mathrm{N2}$ was not observed in our crystal. The sample shows zero-field magnetization below $T_\mathrm{N}$ (see Supplemental Material \cite{Supplemental}), similarly to double-transition samples below $T_\mathrm{N2}$ \cite{Park2022, Park2023, Takagi2023}, although the values in our crystals are an order of magnitude lower, and no AHE is observed. Interestingly, despite the EDX results, magnetic and transport measurements (see Supplemental Material \cite{Supplemental}) suggest slight overdoping, according to Park \textit{et~al.} \cite{Park2024}. Temperature-dependent electrical resistivity shows a kink at the magnetic transition temperature. The residual resistivity ratio (ratio of room-temperature resistivity to resistivity at 2~K) in our crystals was around 2.2, slightly higher than that reported in double-transition samples \cite{Park2022}, indicating high sample quality. Analysis of X-ray diffraction data shows that the compound crystallizes in the hexagonal space group No.~182 ($P6_3 22$), with lattice constants $a = 5.76$~\AA{} and $c = 11.99$~\AA{}.

We also synthesized a single crystal with 22\% Co, as indicated by EDX. For this sample, the starting material contained 33\% Co per Ta. The electrical resistivity exhibited a nonmetallic-like trend, varying only weakly with temperature. Magnetic susceptibility reveals a ferromagnetic-like transition around 25~K. This behavior in the 22\% intercalated sample is consistent with previous reports \cite{Liu2022}.

ARPES measurements were performed at the APE-LE beamline \cite{Panaccione2009} of the Elettra synchrotron. The crystals were oriented along the high-symmetry $\Gamma$M$_0$ direction using Laue diffraction and mounted on copper plates with silver epoxy. Clean surfaces were prepared by cleaving inside the ARPES main chamber under ultra-high vacuum (base pressure $3 \times 10^{-10}$~mbar) at 20~K. The incident photon energy ($h\nu$) was varied between 30 and 90~eV. Most measurements were carried out at 20~K, below the magnetic ordering temperature of Co$_{1/3}$TaS$_2$. To compare spectra below and above the magnetic transition, additional scans along $\Gamma$M$_0$ and Fermi surface maps were recorded at 50~K. The photon beam was linearly polarized in the plane of the storage ring and oriented perpendicular to the analyzer slit. Photoelectrons were collected using a hemispherical DA30 analyzer equipped with a multi-channel plate and CCD detector. The overall energy resolution was approximately 30~meV.

The electronic band structure of Co$_{1/3}$TaS$_2$ was calculated using DFT+U as implemented in the Quantum ESPRESSO package \cite{Giannozzi2009, Giannozzi2017}. We used a kinetic energy cut-off of 120 Ry for the plane-wave basis set and 1000 Ry for the charge density and potentials.
To account for electronic correlations on the cobalt ions, we employed the DFT+U method following the approach of Dudarev \textit{et al.} \cite{Dudarev1998}. The on-site Hubbard interaction parameter, $U = 5.77$ eV, was computed using a linear-response formulation of constrained DFT (LR-cDFT), implemented within the framework of density-functional perturbation theory (DFPT) \cite{Timrov2018, Timrov2021, Timrov2022}.
The Brillouin zone was sampled using a 10$\times$10$\times$5 $k$-point mesh without shift. The sharp cutoff of electron occupation at Fermi energy is modeled by a smearing technique proposed by Marzari and Vanderbilt \cite{Marzari1999}, with a broadening width of 0.01 Ry.


\begin{figure*}[t!] 
\includegraphics[width=0.8\textwidth]{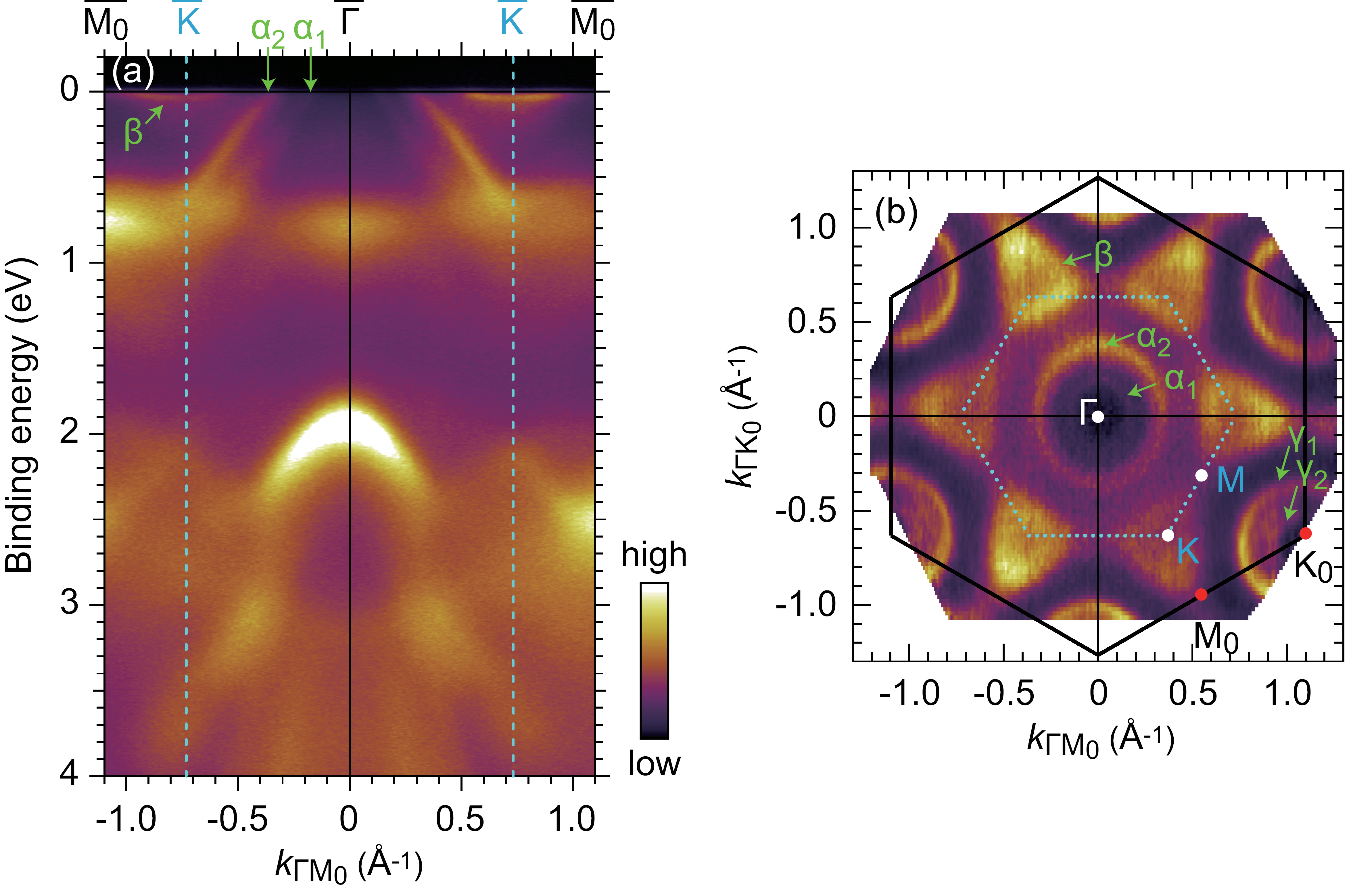}
\caption{(Color online) (a) ARPES intensity plot of Co$_{1/3}$TaS$_2$ along the $\overline{\Gamma}{\overline{\mathrm{M}}}_0$ direction at 20~K measured with $h\nu = 72$~eV. The vertical dashed lines indicate the boundaries of the Co$_{1/3}$TaS$_2$ surface Brillouin zone. (b) Fermi surface map at 20~K measured with $h\nu = 72$~eV. The large black hexagon, drawn with a solid line, represents the first Brillouin zone of 2H-TaS$_2$, and the smaller blue hexagon, shown with dashed lines, represents the first Brillouin zone of Co$_{1/3}$TaS$_2$.}
\label{fig1} \end{figure*}

Electronic correlations were investigated using CPT based on a Wannier-function representation of the spin-polarized DFT+U band structure. The calculations were performed within an energy window of $\pm 1$~eV around the Fermi level. Each cluster consisted of four Wannier nodes representing three Ta and one Co orbitals: the Ta-derived Wannier functions are centered in the triangular plaquettes between Ta atoms, whereas the Co Wannier functions are atom-centered. Local electron interactions on the Co orbitals were included via a Hubbard $U$ term, with double counting corrected by adjusting the on-site energies. The many-body cluster Hamiltonian was solved exactly, while inter-cluster hoppings were treated perturbatively to construct the full CPT Green's function. The resulting spectral functions were unfolded to the extended Brillouin zone for direct comparison with ARPES measurements. 
\section{Results and discussions}

Fig.~\ref{fig1} shows the ARPES intensity plots of Co$_{1/3}$TaS$_2$ measured at 20~K with $h\nu = 72$~eV. Panel (a) presents spectra collected along the $\overline{\Gamma}\overline{\mathrm{M}}_0$ direction, where $\overline{\mathrm{M}}_0$ corresponds to the high-symmetry point in the surface Brillouin zone (BZ) of the host 2H-TaS$_2$, while $\overline{\mathrm{K}}$ marks the high-symmetry point of the Co$_{1/3}$TaS$_2$ superlattice BZ.

The main band features resemble those of pristine 2H-TaS$_2$ \cite{Camerano2025}, with no obvious band folding or splitting at the boundaries of the Co$_{1/3}$TaS$_2$ BZ. Beyond an overall band shift due to charge transfer, a slight redistribution of Ta and S atoms upon Co intercalation introduces a weak structural modulation of the host lattice. This modulation, together with occupational disorder of intercalated Co atoms, would in principle lead to band backfolding and splitting in the reconstructed Brillouin zone; however, due to its small amplitude these features are not resolved in ARPES and manifest instead as an apparent broadening of the TaS$_2$-derived bands.

Fig.~\ref{fig1}(b) shows the corresponding Fermi surface map. The large black hexagon and the smaller blue dashed hexagon indicate the surface Brillouin zones of 2H-TaS$_2$ and Co$_{1/3}$TaS$_2$, respectively.
As in 2H-TaS$_2$, double-walled, cylindrical Fermi surfaces are observed around the $\overline{\Gamma}$ point ($\alpha_1$ and $\alpha_2$) and the $\overline{\mathrm{K}}_0$ point ($\gamma_1$ and $\gamma_2$) \cite{Zhao2017, Wijayaratne2017}.
Compared to the parent compound, in Co$_{1/3}$TaS$_2$, the splitting between these double walls is more pronounced around the $\overline{\Gamma}$ point due to band shifting, and more importantly, Co-mediated interlayer hybridization. The same effect was also observed in Co$_{1/3}$NbS$_2$ \cite{Popcevic2022}.
The inner FS ($\alpha_1$) is very faint, representing a bonding band. Additionally, a notable difference introduced by intercalation is the emergence of a triangular Fermi surface pocket, commonly referred to as the "$\beta$ feature", centered at the $\overline{\mathrm{K}}$ point. The band responsible for this Fermi surface pocket is a shallow, electron-like band around the $\overline{\mathrm{K}}$ point, which is marked by the blue dashed line in Fig.~\ref{fig1}(a). Similar features have been reported in other intercalated dichalcogenides such as Co$_{1/3}$NbS$_2$ \cite{Popcevic2022, Tanaka2022, Yang2022}, Ni$_{1/3}$NbS$_2$ \cite{Utsumi2024}, and Cr$_{1/3}$NbS$_2$ \cite{Sirica2016, Sirica2020}.
The observation of the $\beta$ feature in Co$_{1/3}$TaS$_2$ demonstrates that this shallow electron band is not limited to NbS$_2$-based compounds, suggesting a more general, intercalant-driven effect in transition metal dichalcogenides.

To study the possible effect of magnetic ordering on the band structure, we also measured the ARPES spectra and the Fermi surface above the magnetic transition ($T_\mathrm{N} = 37$~K) at 50~K. There was no observable difference between the spectra at 20~K and 50~K (see Supplemental Material \cite{Supplemental}). The presence of the $\beta$ feature above and below $T_\mathrm{N}$ confirms that the origin of this shallow electron pocket is independent of the magnetic transition. Similarly, the antiferromagnetic TMDs Co$_{1/3}$NbS$_2$ and Ni$_{1/3}$NbS$_2$ also showed no detectable changes in the electronic structure across their magnetic transitions \cite{Popcevic2023, Utsumi2024}.

Although the $\beta$ feature has been experimentally observed in NbS$_2$ intercalates, its origin remains debated. One possible explanation, invoking a surface state, has been proposed primarily because in Co$_{1/3}$NbS$_2$ its relative intensity varies with surface termination or position \cite{Tanaka2022}. Nevertheless, its dependence on photon energy and observation in soft X-ray ARPES \cite{Yang2022} do not convincingly support this scenario. Furthermore, our DFT calculations for a two-layer slab of Co$_{1/3}$TaS$_2$ do not reproduce the $\beta$ feature as a surface state, indicating that it is unlikely to originate from surface effects (see Supplemental Material \cite{Supplemental} see also references \cite{Kokalj1999, Kokalj2003}).

Apart from the $\beta$ feature, the DFT calculations successfully reproduce the experimentally observed band structures of Co$_{1/3}$NbS$_2$ \cite{Popcevic2023} and Ni$_{1/3}$NbS$_2$ \cite{Utsumi2024}. In particular, the DFT bands, when unfolded into the larger parent-compound Brillouin zone, match the main band features observed by ARPES. The same is true for Co$_{1/3}$TaS$_2$ (see Appendix~B).


\begin{figure*}[t!] 
\includegraphics[width=\textwidth]{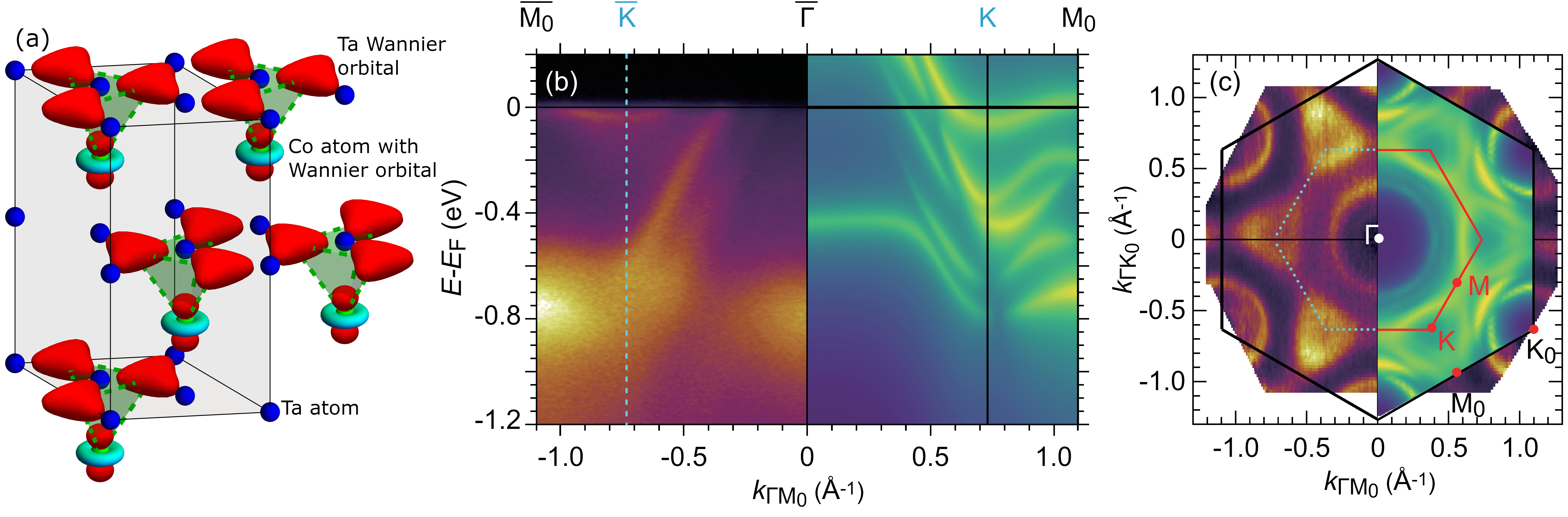}
\caption{(Color online) (a) Graphic representation of the crystal structure. Co atoms are shown as green spheres, overlaid with the corresponding Co Wannier orbitals. Red isosurfaces represent the interstitial Ta Wannier orbital lobes located between neighboring Ta atoms rather than at the atomic positions themselves (shown as blue spheres). Intra-cluster connections are indicated by dashed green lines and the clusters are shaded in green. The unit cell is indicated by a thin black line and shaded in gray. Sulfur atoms coordinating Ta in trigonal prismatic arrangement are omitted for clarity. These sulfur atoms occupy positions such that the edges of the trigonal prisms, connecting sulfur atoms in the layers above and below, do not intersect the Ta Wannier orbitals; that is, the sulfur triangles are rotated by $30^\circ$ with respect to the triangles defined by the Wannier orbital positions. Comparisons of the spectra presented in Fig.~\ref{fig1} with the results of the CPT calculation: (b) the band structure along $\overline{\mathrm{\Gamma}}{\overline{\mathrm{M}}}_0$ direction and (c) the Fermi surface map of Co$_{1/3}$TaS$_2$.}
\label{fig2} 
\end{figure*}

We attribute the origin of the $\beta$ feature to strong correlation effects that cannot be captured by standard DFT+U. In our earlier study of Co$_{1/3}$NbS$_2$ \cite{Popcevic2023}, the feature was addressed within a slave-boson approach, where it was shown that strong local Coulomb interactions on the Co sites lead to a renormalization of the Co-host hybridization and a shift of the Co-derived states toward the Fermi level, resulting in the emergence of a narrow, resonance-like band near the $\overline{\mathrm{K}}$ point, consistent with the experimentally observed $\beta$ feature. The strong-correlation origin of this feature was also previously examined within the dynamical mean-field theory (DMFT) framework \cite{Park2024b}, yielding partial agreement with experiment.

While these analyses provided qualitative evidence for the correlation-driven origin of the $\beta$ feature, in the present work we adopt a more exact treatment of electronic correlations by employing CPT, initially proposed by Gros \textit{et al.}~\cite{Gros1993} and further developed by S n chal \textit{et al.}~\cite{Senechal2000, Senechal2002}. In CPT, the Hubbard interaction is treated exactly at the cluster level, going beyond the mean-field approximation inherent to DFT+U and capturing momentum-resolved spectral features that are not fully captured by DMFT.

Our CPT calculation is based on a Wannier function description of the band structure obtained from DFT+U for the spin-polarized system. We focus on the energy range around the Fermi level where the "$\beta$ feature" appears, in particular, from 1.0 eV below to 1.5 eV above the Fermi level. In this energy range, seven DFT bands can be accurately fitted (for details see lower graph in Fig.~\ref{fig:wannier90}, shown in Appendix~A) with a tight-binding type Hamiltonian obtained from the Wannier90 package~\cite{Pizzi2020} using seven Wannier functions. Several of the most dominant terms of the Wannier Hamiltonian are listed in Table~I. While both spin channels share the same band structure, the Wannier representation of DFT results is spin-dependent. The six Wannier functions (spin-independent) are predominantly of Ta $d$-state character and are located in the space forming a triangle between three Ta atoms, not directly at the Ta atom positions (see Fig.~\ref{fig2}(a)). The seventh Wannier function is spin-dependent: for electrons with spin up, its location is at the Co atom with spin down, and vice versa. This Wannier function primarily describes the unoccupied band above the Fermi level, while the Ta-derived Wannier functions capture the dispersions of the six bands crossing the Fermi level.

The CPT calculation requires a unified Hamiltonian that incorporates both spin projections. This is achieved by merging the Wannier Hamiltonians for the two spins, resulting in a Hamiltonian with eight Wannier functions per spin and per unit cell (shown in Fig.~\ref{fig2}(a)). Two of these Wannier functions are located on Co atoms. To account for electron correlations, it is necessary to include the Hubbard interaction on the Co sites into the full CPT Hamiltonian. However, the single-particle part of the full Hamiltonian, obtained from DFT+U, already incorporates a mean-field approximation of the Hubbard interaction, and we must correct for double counting.

Double counting is a well-known and subtle problem that appears in all theories based on Wannier Hamiltonians going beyond a simple mean-field approximation, e.g., DMFT. It is extensively studied in many works~\cite{Ryee2018a, Ryee2018b, Karolak2010, Park2014, Nekrasov2012, Kotliar2006, Katsnelson2008, Haule2015, Plekhanov2018}. Here, we adopt the simplest approach, which is exact only in the localized limit: the double counting is corrected by modifying the Wannier on-site energies of sites with Hubbard interaction. Assuming the Hubbard interaction is given by:
\begin{equation}
H_\mathrm{Hub} = U \cdot ( n_\uparrow - 0.5 ) ( n_\downarrow - 0.5 ),
\end{equation}the Wannier on-site energy for unoccupied sites should be reduced by $0.5U$, and the Wannier on-site energy for occupied sites should increase by $0.5U$. In our case, Co Wannier functions/sites are unoccupied; therefore, their on-site energy should be reduced by $0.5U$.

The starting point of CPT is disconnected clusters with a finite number of sites tiling the entire crystal lattice. The many-body Hamiltonian of these isolated clusters can be solved exactly, either by exact diagonalization or other appropriate methods, e.g., Monte Carlo simulation. In the next step, the neglected parts of the full Hamiltonian, i.e., the hopping terms between clusters, are treated perturbatively. This perturbative part can be accounted for exactly to all orders. Assuming that the Green's function of the system of disconnected clusters is $G_0(\omega)$, which is an infinite matrix in the site representation, the CPT Green's function is:
\begin{equation}
G(\omega) = \left[ G_0(\omega)^{-1} - V \right]^{-1},
\end{equation}
where $V$ is the part of the full Hamiltonian treated perturbatively. The problem of the infinite matrices is resolved using a Fourier transformation since all clusters form a periodic lattice themselves. One should note that even though both steps are exact, the final CPT Green's function is approximate; some many-body contributions are not fully accounted for. Nevertheless, this method represents a significant improvement in treating many-body correlations.

In our CPT model, the clusters are built from three interstitial Ta-character Wannier functions and one Co-centered Wannier function per spin. The Co$_{1/3}$TaS$_2$ unit cell contains two clusters, as shown in Fig.~\ref{fig2}(a). The Green's functions are calculated for real frequencies and at finite temperatures (300~K) using the Lehmann representation. The chemical potential is chosen to fix the average number of particles in the cluster and in the system. The calculated Green's function has wave-number periodicity corresponding to the Co$_{1/3}$TaS$_2$ unit cell, i.e., the small Brillouin zone indicated by the red line in Fig.~\ref{fig2}(c). The Green's function in the extended Brillouin zone (band unfolding) is obtained using Senechal's G-scheme (Green's function periodization) within CPT
~\cite{Senechal2012}.


\begin{figure}[t!] 
\includegraphics[width=0.5\textwidth]{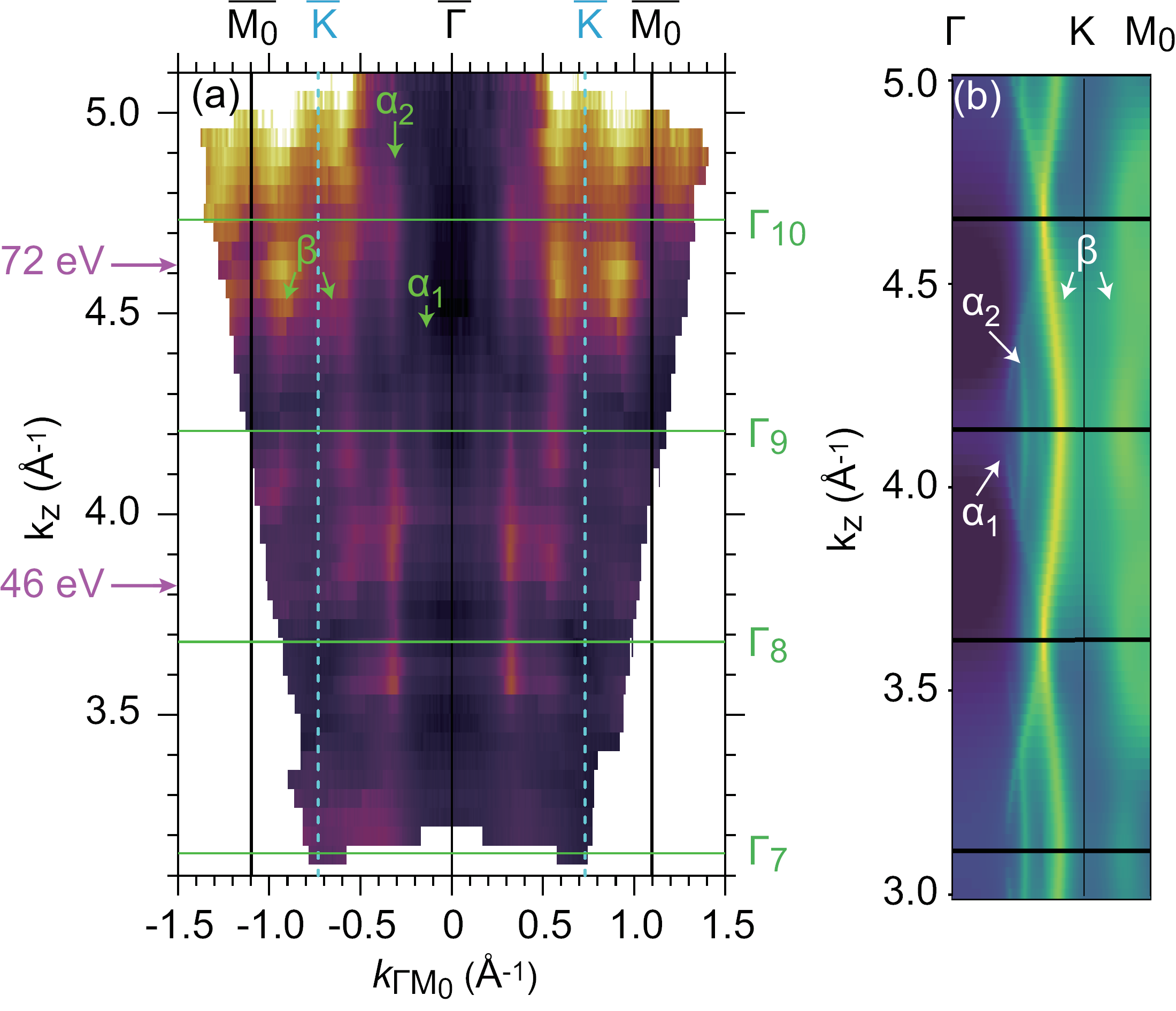}
\caption{(Color online) (a) ARPES intensity plot along the $k_z$ direction at the Fermi level. The vertical dashed lines correspond to the boundaries of the Co$_{1/3}$TaS$_2$ Brillouin zone. The horizontal green lines and labels ($\mathrm{\Gamma} _{n+1}$) mark the $k_z$ levels corresponding to  $\mathrm{\Gamma}_{n+1}$ $\equiv$ (0, 0, 2$\pi/c \times n$). (b) The CPT calculated Fermi surface along $k_z$-direction.
}
\label{fig3} 
\end{figure}

The results of the CPT calculations, expressed as the spectral function $A(\vec{k},\omega) = -\frac{1}{\pi} \, \mathrm{Im} \, G(\vec{k},\omega)$, are presented and compared to the experimental band structure and the Fermi surface collected with $h\nu = 72$~eV at 20~K in Fig.~\ref{fig2}(b) and (c). Keeping in mind the simplicity of our model and the fact that ARPES intensities are further modulated by matrix element effects and finite experimental resolution we can see notable resemblance between experimental and calculated spectra.

The most prominent difference from the standard DFT+U results (for details see Fig.~\ref{fig:DFTunfold} shown in Appendix~B) is the emergence of a shallow electron-like band with its minimum located at the ${\overline{\mathrm{K}}}$ point, clearly visible on the right side in Fig.~\ref{fig2}(b). In the corresponding Fermi surface map (Fig.~\ref{fig2}(c)), this band forms a triangular contour around the ${\overline{\mathrm{K}}}$ point, closely reproducing the experimentally observed $\beta$ feature. The orbital character analysis indicates that this state has a substantial Co contribution. Its appearance only within the CPT framework, which treats the Hubbard $U$ beyond the mean-field level, provides strong evidence that the $\beta$ feature originates from strong electronic correlations, in line with our earlier conjecture. In contrast, the remaining bands in the CPT calculation are very similar to those from DFT+U, which already provides a satisfactory description of the corresponding experimental spectra. An exception is the spectral weight around 0.8 eV binding energy, which originates from the highest S-derived band (see Appendix B) and is not captured by our reduced CPT model.


\begin{figure*}[t!] \includegraphics[width=\textwidth]{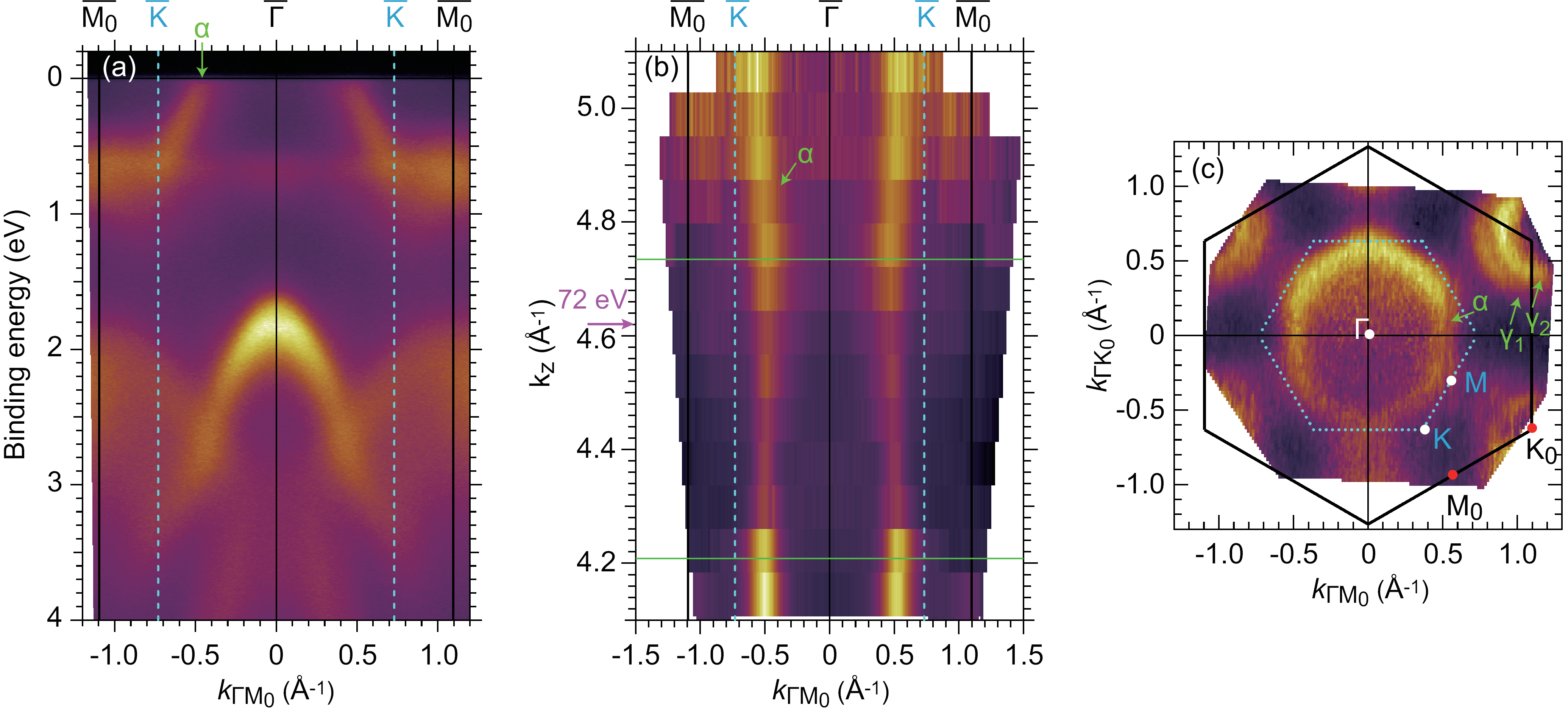}
\caption{(Color online) (a) ARPES intensity plot of Co$_{0.22}$TaS$_2$ measured at $h\nu = 72$~eV and 20~K. (b) Fermi surface along the $k_z$ direction, obtained by varying the photon energy from 57 to 90~eV in 3~eV steps at 20~K. (c) Fermi surface measured at $h\nu = 63$~eV and 20~K.
}

\label{fig4} 
\end{figure*}

It is important to stress that the present calculation is intentionally minimalistic, retaining only a limited set of bands near the Fermi level. Consequently, the residual discrepancies between CPT and experiment may arise from several sources: (i) the intrinsic approximations of the CPT method, (ii) the reduced orbital and spin manifold employed in the model, which effectively describes the Co sites by a single spin-$1/2$ orbital, whereas the real Co ions are in a high-spin $S = 3/2$ configuration that requires the explicit inclusion of additional Co $d$ orbitals and a proper treatment of Hund's rule coupling, (iii) the simplified treatment of the double-counting correction, and (iv) the neglect of direct hopping between intercalated Co atoms with opposite spin polarization. Additional factors, such as finite experimental resolution and the artificial broadening applied to the calculated spectra, are also expected to contribute.

Fig.~\ref{fig3} (a) shows the Fermi surface map of Co$_{1/3}$TaS$_2$ along the $k_z$ direction, obtained by varying the photon energy from 30 to 90 eV in small steps. The photon energies corresponding to the measurements shown in Fig.~\ref{fig1} and Fig.~\ref{fig:DFTunfold}(a) are indicated. The data were processed using the same procedure as described in Refs.~\cite{Popcevic2022, Utsumi2024}. The $k_z$ dispersion plot reveals a two-dimensional (2D-like) Fermi surface, $\alpha_2$, visible at $k_x \approx \pm 0.3\text{\AA}^{-1}$, which does not disperse with $k_z$ and shows only intensity variations. A faint additional 2D-like Fermi surface, $\alpha_1$, can also be observed near $k_x \approx \pm 0.15\text{\AA}^{-1}$, although it is too weak to be traced across all $k_z$ values.

Both features arise from the Ta $d$ bands of 2H-TaS$_2$: the outer band ($\alpha_2$) corresponds to the antibonding state, while the inner ($\alpha_1$), weaker one corresponds to the bonding state. DFT modeling predicts that the bonding band should exhibit noticeable $k_z$ dispersion and cross the Fermi level between the $\Gamma$ and A points. This explains why the $\alpha_1$ band is not observed across all $k_z$ values in Fig.~\ref{fig3}(a). Indeed, the ARPES intensity plots measured with photon energies between $h\nu = 38$-50~eV (for example see Fig.~\ref{fig:DFTunfold}(a), presented in Appendix~B) show the bonding band ($\alpha_1$) below the Fermi level. This photon energy range corresponds to $k_z$ values below 4~$\text{\AA}^{-1}$ in Fig.~\ref{fig3}(a), which accounts for the disappearance of $\alpha_1$ at low $k_z$ values.

This $k_z$ dispersion agrees with the previously observed enhancement of $c$-axis electronic transport in TMDs upon Co intercalation, as discussed for Co$_{1/3}$NbS$_2$ in Refs.~\cite{Popcevic2022, Popcevic2023}. The $\beta$ feature appears in Fig.~\ref{fig3}(a) on both sides of the $\overline{\mathrm{K}}$ point, with its intensity increasing at higher photon energies, consistent with the behavior seen in Co$_{1/3}$NbS$_2$.

Fig.~\ref{fig3}(b) shows the same Fermi surface as in panel (a), but calculated using the CPT method. The $\beta$ feature is again visible on both sides of the K point. Ta $d$ bands are also visible with antibonding band showing virtually no dispersion while bonding band exhibits strong $k_z$ dependence. We also notice the characteristic periodicity of two Brillouin zones resulting from the fact that there are two Co and two Ta layers per one unit cell. A similar effect was already reported by Weber \textit{et al.}~\cite{Weber2018} in NbSe$_2$.

To investigate the effect of reduced Co concentration on the electronic structure, ARPES measurements were performed on the underdoped Co$_{0.22}$TaS$_2$. The results are shown in Fig.~\ref{fig4}. Notably, the band structure and Fermi surface closely resemble a rigid-band-shifted 2H-TaS$_2$ structure. By varying the incident photon energy from 57 to 90~eV, in Fig.~\ref{fig4}(b) we mapped the Fermi surface along the $k_z$ direction. It exhibits a quasi-2D character with only a small modulation, and no $\beta$ feature is observed within the experimental photon energy range.

In Fig.~\ref{fig4}(c), the Fermi surface shows the central $\alpha$ pocket around the $\mathrm{\Gamma}$ point and the $\gamma_1$ and $\gamma_2$ pockets around K$_0$, while the $\beta$ feature, visible in Fig.~1 around K, is not observed. The $\alpha$ pocket lies at $k_x \sim 0.5~\text{\AA}^{-1}$, very close to its position in 2H-TaS$_2$ \cite{Camerano2025}, indicating a much smaller band shift compared to the 33\% intercalated compound and suggesting even lower charge transfer.

The same indication comes from the magnetic moment of Co$_{1/3}$TaS$_2$ compared to Co$_{0.22}$TaS$_2$. 
Co$_{1/3}$TaS$_2$ exhibits an effective magnetic moment of $\sim 3.3\,\mu_B$ similar to Co$_{1/3}$NbS$_2$ \cite{Popcevic2023}.
Despite being reduced compared to the spin-only value
$\mu_\mathrm{eff} = 3.87\,\mu_B$ for $S=3/2$, this moment is fully
consistent with a $\mathrm{Co}^{2+}$ configuration once
hybridization with the conduction band is taken into account, as
demonstrated by DFT calculations \cite{Popcevic2023, Utsumi2024}.
The underdoped compound Co$_{0.22}$TaS$_2$ exhibits an effective
magnetic moment of $\sim 2.6\,\mu_B$ obtained from the Curie-Weiss fit,
close to the spin-only value $2.83\,\mu_B$ for $S=1$, indicating a
reduced effective spin state, consistent with a lower charge transfer.
The lower charge transfer in underdoped compound implies that the lowest unoccupied Co orbital taken into account in our CPT is now below Fermi level, leaving only two remaining Co $d$ orbitals above the Fermi level. This influences the energy spectrum, making it more similar to the Ni intercalated case where $\beta$ feature was much weaker and shallower\cite{Utsumi2024}. 
Furthermore, the disorder inevitably present in Co$_{0.22}$TaS$_2$, as evidenced by the negative temperature dependence of the electrical resistivity and relatively broad ferromagnetic onset, predominantly affects the Co-derived electronic states, suppressing the coherence required for the formation of a well-defined $\beta$ feature. The remaining bands are TaS$_2$-derived and are much less affected by disorder, retaining sufficient coherence to remain observable in ARPES, although they appear significantly broadened.
A difference is also seen in the shape of the central Fermi surface presented in Fig.~\ref{fig4}(c), which in Co$_{0.22}$TaS$_2$ exhibits a hexagonal form, similar to that observed in Ni$_{1/3}$NbS$_2$ \cite{Utsumi2024}, in contrast to the circular Fermi surfaces of 33\% Co intercalates~\cite{Popcevic2022, Tanaka2022}.

Furthermore, the disorder inevitably present in Co$_{0.22}$TaS$_2$, as evidenced by the negative temperature dependence of the electrical resistivity and relatively broad ferromagnetic onset (\cite{Liu2022}, see also Supplemental Material \cite{Supplemental}), predominantly affects the Co-derived electronic states, suppressing the coherence required for the formation of a well-defined $\beta$ feature. This also explains the absence of the inner $\alpha_1$ band, which has a significant Co orbital character \cite{Popcevic2023} and is therefore particularly sensitive to disorder within the Co subsystem. The remaining bands are TaS$_2$-derived and are much less affected by disorder, retaining sufficient coherence to remain observable in ARPES, although they appear significantly broadened as a result of disorder-induced local lattice distortions around intercalated Co ions.

A difference is also seen in the shape of the central Fermi surface presented in Fig.~\ref{fig4}(c), which in Co$_{0.22}$TaS$_2$ exhibits a hexagonal form, similar to that observed in Ni$_{1/3}$NbS$_2$ \cite{Utsumi2024}, in contrast to the circular Fermi surfaces of 33\% Co intercalates~\cite{Popcevic2022, Tanaka2022}.


\section{Conclusion}

We have investigated the electronic structure and Fermi surface of Co$_{1/3}$TaS$_2$ using ARPES combined with advanced theoretical modeling of electronic correlations. In this 33\% intercalated compound, a shallow electron pocket, the so-called $\beta$ feature, emerges as a clear deviation from a rigid-band shift, while it is notably absent in the underdoped Co$_{0.22}$TaS$_2$, highlighting the sensitivity of this feature to the intercalant concentration. To our knowledge, this represents the first experimental observation of the $\beta$ feature in an intercalated TaS$_2$ system. Unlike previous interpretations in Co$_{1/3}$NbS$_2$, where this feature was ascribed to surface effects, our CPT modeling of Co$_{1/3}$TaS$_2$ reproduces the $\beta$ feature and demonstrates that it originates from strong electronic correlations, strongly influenced by the intercalated cobalt orbitals.

These results build on our earlier indications in Co$_{1/3}$NbS$_2$, where tight-binding modeling of DFT results combined with a slave-boson approach suggested a correlation-driven origin of the $\beta$ feature. Independently, DMFT calculations reported in the literature \cite{Park2024b} pointed in a similar direction. In the present study, by employing CPT, we provide the first quantitative modeling that reproduces the experimental ARPES spectra of Co$_{1/3}$TaS$_2$, thereby directly demonstrating that strong electronic correlations are essential for understanding the low-energy electronic structure of intercalated transition-metal dichalcogenides.

Our study further demonstrates that the $\beta$ feature depends sensitively on the orbital character of the intercalant and the intrinsic disorder associated with the Co ions. These results underscore the necessity of incorporating strong correlation effects in future theoretical studies of intercalated transition-metal dichalcogenides.

\section*{Acknowledgements}
       We acknowledge insightful discussions with Eduard Tuti\v{s}. This work has been partly supported by the Croatian Science Foundation under the project numbers IP-2020-02-9666, UIP-2019-04-2154 and IP-2022-10-3382, by the project: Ground states in competition - strong correlations, frustration and disorder - FrustKor financed by the Croatian Government and the European Union through the National Recovery and Resilience Plan 2021-2026 (NPOO) and by the project Cryogenic Centre at the Institute of Physics - KaCIF co-financed by the Croatian Government and the European Union through the European Regional Development Fund-Competitiveness and Cohesion Operational Programme (Grant No. KK.01.1.1.02.0012). The research has been performed at the Elettra at the APE-LE beamline (proposal No. 20235254). NB acknowledges support by the project CeNIKS co-financed by the Croatian Government and the European Union through the European Regional Development Fund Competitiveness and Cohesion Operational Program (Grant No. KK.01.1.1.02.0013). The work at TU Wien was supported by the Austrian Science Fund (FWF) [10.55776/F86; 10.55776/P35945]. This work has been partly performed in the framework of the nanoscience foundry and fine analysis (NFFA-MUR Italy Progetti Internazionali) facility. The research leading to this result has been co-funded by the project NEPHEWS under Grant Agreement No 101131414 from the EU Framework Programme for Research and Innovation Horizon Europe

\begin{figure}[b!]
\centering
\includegraphics[width=0.45\textwidth]{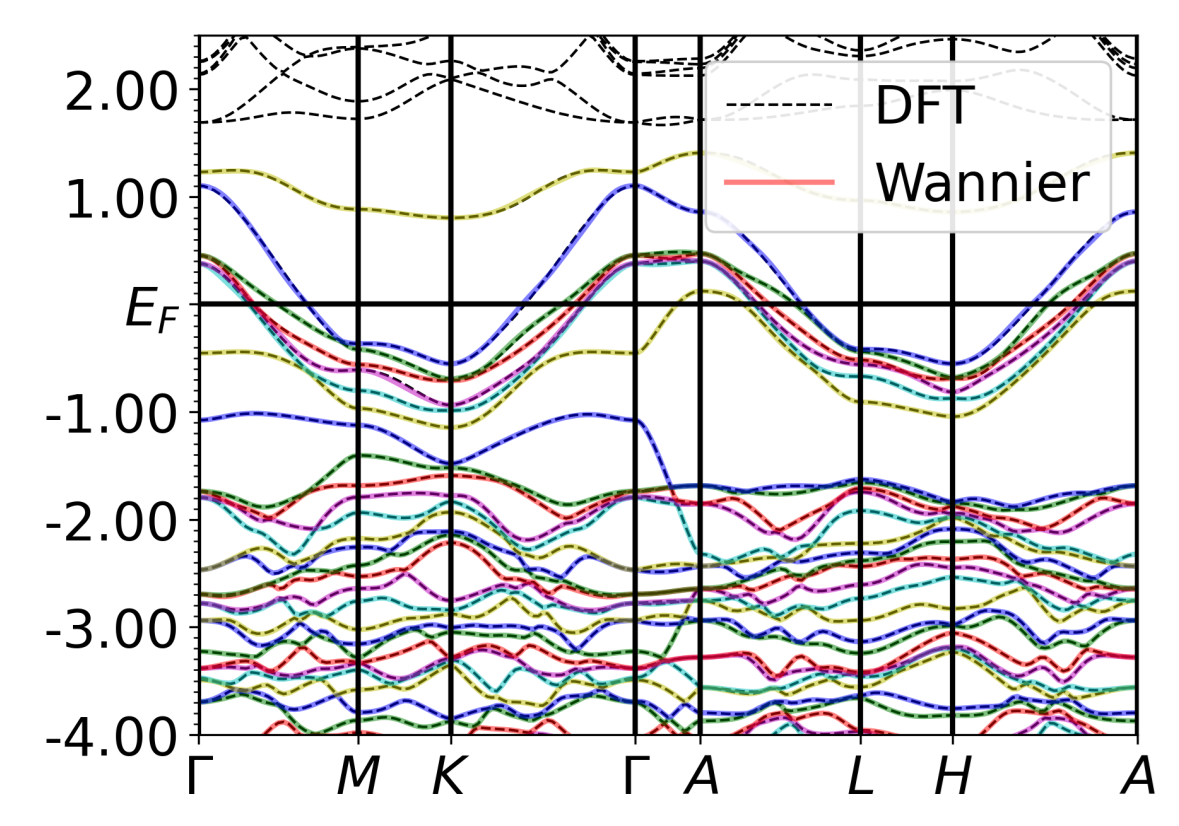}
\includegraphics[width=0.45\textwidth]{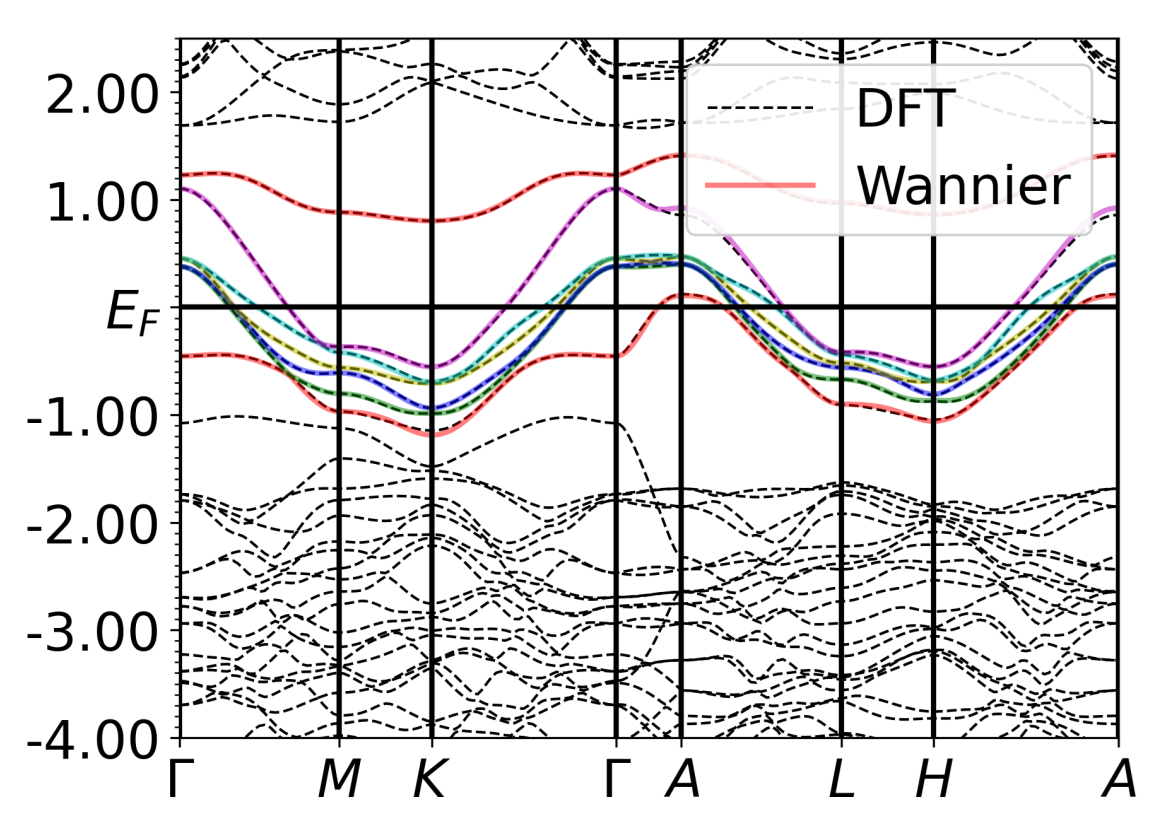}
\caption{(Color online) Comparison of the DFT band structure (dotted lines) with the band structures obtained from the Wannier Hamiltonians, (solid lines). The upper panel shows the band tructure constructed using 52 Wannier functions per unit cell, while the lower panel displays the result obtained with a reduced basis of 7 Wannier functions per unit cell.}

\label{fig:wannier90}
\end{figure}

\section* {Appendix A. Wannier functions and Hamiltonian}

The idea of using Wannier functions is to develop a Hubbard like
model in which the Hubbard interaction on the intercalated atoms
will be treated more accurately compared to DFT calculations.
The Wannier functions can be extracted from DFT by using software
package Wannier90 \cite{Pizzi2020}. We have performed two versions of Wannier calculations. The first one, focused on the energy range from the Fermi level ($E_{\mathrm{F}}$) - 7.7 eV to $E_{\mathrm{F}}$ + 1.5 eV, is based on 52 Wannier functions within the unit cell of Co$_{1/3}$TaS$_2$. The space localized nature of Wannier functions makes possible to
create a tight-binding like Hamiltonian which can accurately reproduce
DFT band structure within given energy range. The result is shown
in Fig.~\ref{fig:wannier90} in the upper panel.

The second, simplified version of the Wannier calculation, is developed to describe the energy range just around the Fermi level, from $E_{\rm F}$-1.0 eV to $E_{\rm F}$+1.5 eV as this is the energy range where $\beta$ feature is observed. It is based on 7 Wannier functions in the unit cell.
One may see that corresponding Hamiltonian can accurately reproduce 7 DFT bands around Fermi level, (the lower panel of Fig.~\ref{fig:wannier90}). The top most Wannier band that is colored with red and is entirely above the Fermi level, has the largest contribution from lowest in energy empty Co $d$-state. The center of this Wannier function is located on Co atom. This Wannier function can be identified as an empty upper Hubbard state while the lower Hubbard state, that is occupied with an electron, is much lower in energy (around 6 eV) and it is outside the energy range in consideration. If this is a dispersion from spin up electrons, the center of Wannier function is located on the spin down Co ion, and \textit{vice versa}. The centers of these Wannier functions depend on the electron spin in consideration. In spite of this differences between different spins, the band spectra for both spin projections are alike. The remaining 6 lower in energy bands have the largest contribution from Ta $d$-states. These Ta Wannier functions are not located on the Ta atoms, but in the points between three Ta atoms, in the space points not surrounded with sulphur atoms from above and below. These Wannier functions are shared with three neighboring Ta atoms. Their positions are independent on the spin projection. The most prominent terms of the simplified Wannier Hamiltonian are listed in Table.\ref{tab:h7w-params}. The largest hopping term in the Wannier Hamiltonian is between Co Wannier and Ta Wannier functions.

\begin{table}[H]
 \centering
 \begin{tabular}{ |l||r| }
 \hline
 parameters & (eV) ~~~ \\
 \hline\hline
 $t$(\text{Co-Ta})   &  0.263293 \\
 \hline
 $t$(\text{Ta-Ta})   &  0.064378 \\
 \hline
 $t^{(2nn)}$(\text{Ta-Ta}) &  0.132622 \\
 \hline
 $t$(\text{Co-Co})   &  0.012778 \\
 \hline
 $e$(\text{Co})           & 12.318694 \\
 \hline
 $e$(\text{Ta})           & 11.412116 \\
 \hline
 \end{tabular}
 \caption{ Dominant terms in the Wannier Hamiltonian. $t$ are the first nearest neighbor hopping between Wannier orbitals, $t^{(2nn)}$ is the second nearest neighbor hopping and $e$ are Wannier orbital's on-site energies. The double counting term is not subtracted. }
 \label{tab:h7w-params}
\end{table}

\section* {Appendix B. DFT+U unfolding}

\begin{figure*}
\centering
\includegraphics[width=\textwidth]{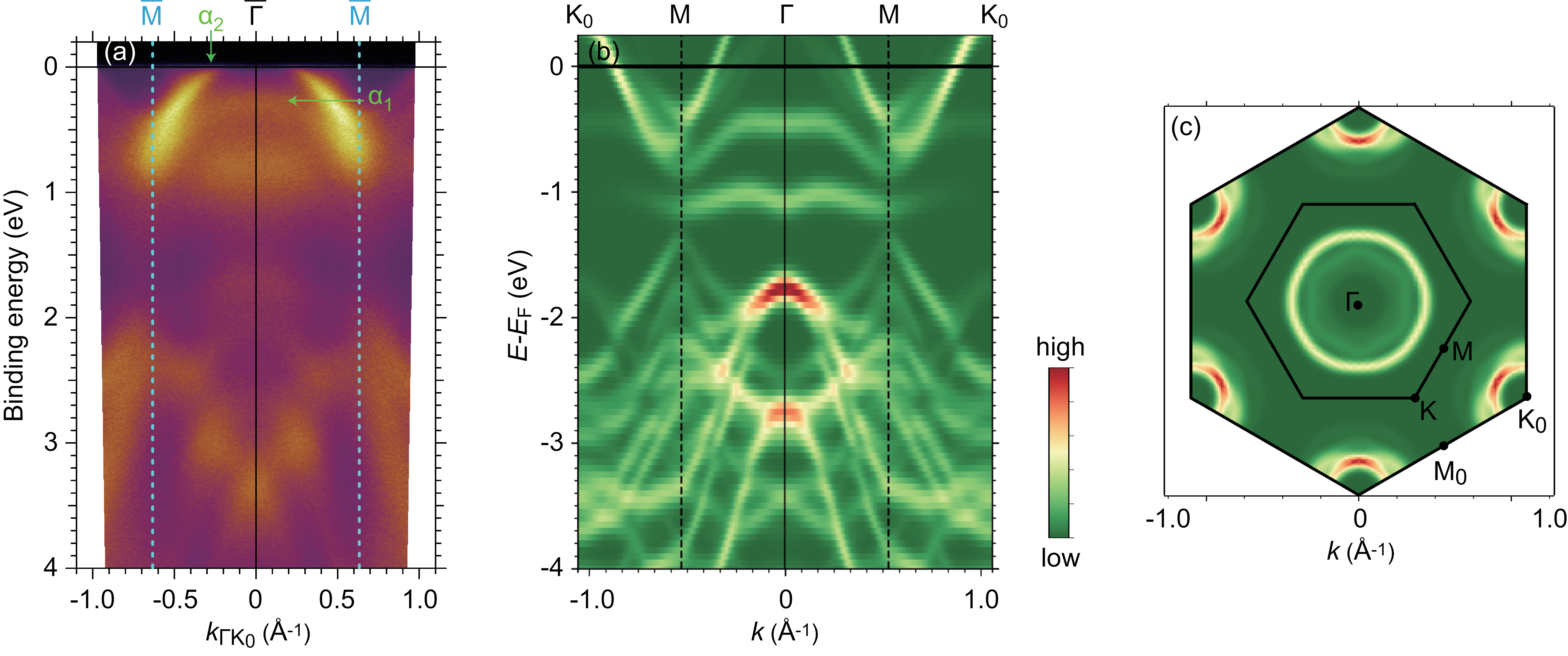}
\caption{(Color online) (a) ARPES spectra measured with h$\nu$=46 eV along the high-symmetry $\mathrm{\Gamma}M$K$_{0}$ direction on Co$_{1/3}$TaS$_2$. (b) DFT calculated band structure of Co$_{1/3}$TaS$_2$ unfolded into the first 2H-TaS$_2$ Brillouin zone using unfold-x package\cite{Popescu2012, Pacile2021}. (c) Fermi surface obtained by unfolding DFT calculated band structure. Inner hexagon represents the BZ of Co$_{1/3}$TaS$_2$, while the outer one represents the BZ of 2H-TaS$_2$.}
\label{fig:DFTunfold}
\end{figure*}

In Fig.~\ref{fig:DFTunfold}(a), the experimental ARPES spectra obtained along the high-symmetry $\mathrm{\Gamma}M$K$_{0}$ measured with h$\nu$=46 eV are shown.
In contrast to Fig.~\ref{fig1}(a), where Ta bonding band is crossing the Fermi level forming the smallest, faint circular Fermi surface around $\mathrm{\Gamma}$ point in Fig.~\ref{fig1}(b), here this band can be seen below Fermi level at the $\mathrm{\Gamma}$ point, together with the highest-lying S-derived band beneath it. In Fig.~\ref{fig:DFTunfold}(b), DFT calculated band structure unfolded into the first Brillouin zone of 2H-NbS$_2$ is shown along the same high symmetry direction. In panel (c), the Fermi surface is shown. The most striking difference between Fig.~\ref{fig1}(b) and Fig.~\ref{fig:DFTunfold}(c) is the absence of $\beta$ feature in DFT calculated Fermi surface.  


%

\end{document}